\documentclass[submission,copyright,creativecommons]{eptcs}
\providecommand{\event}{NCMA 2022} % Name of the event you are submitting to
\begin{document}
\newcommand{\PF}{\noindent {\em \textbf{Proof.} }}
\newcommand{\QED}{\noindent {\em \textbf{QED.} }}
\newcommand{\theo}{\noindent {\em \textbf{Theorem} }}
\newcommand{\lem}{\noindent {\em \textbf{Lemma} }}
\newcommand{\ex}{\noindent {\em \textbf{Example} }}
\newcommand{\cor}{\noindent {\em \textbf{Corollary} }}
\newcommand{\prop}{\noindent {\em \textbf{Proposition} }}

  \title{A Finite-Automaton Based Stream Cipher As a Quasigroup Based Cipher\thanks{This work was supported by United Arab Emirates Program for Advanced Research (UAEU UPAR) Grant No. G00003431.}}
  \author{P\'al D\"om\"osi  
   \institute{Faculty of  Informatics, University of Debrecen\\
   	H-4028 Debrecen, Kassai \'ut 26, Hungary}
   	\institute{Institute of Mathematics and  Informatics,
   	University of  Ny{\'i}regyh\'aza\\
   	H-4400 Ny{\'i}regyh\'aza, S\'ost\'oi \'ut 36, Hungary}
   	\email{domosi@unideb.hu}  
      \and
      Adama Diene  %\\
      \institute{Department of Mathematical Science, United Arab Emirates University,\\
      P.O. Box 15551, Al Ain, Abu Dhabi, United Arab Emirates}
     \email{diene@uaeu.ac.ae}
    }
 % \titleodd{A Finite-Automaton Based Stream Cipher as a Quasigroup Cipher\thanks{This 
 % 		work was supported by United Arab Emirates Program for Advanced Research (UAEU UPAR)  Grant \# G00003431.}
 % }
 % \authoreven{P. \'al D\"om\"osi and A. Diene}
%\keywords{stream cipher,finite automaton, quasigroup}
%%%%%%%%%%%  \begin{document}
 \def\titlerunning{A Finite-Automaton Based Stream Cipher As a Quasigroup Based Cipher}
\def\authorrunning{P\'al D\"om\"osi and Adama Diene}
  	\maketitle
  \begin{abstract}
  	In this paper we show that a recently published finite automaton stream cipher can be considered as a quasigroup based 
  	stream cipher. Some additional properties of the discussed cipher are also given.
  \end{abstract}
 %%% \abstractSi{}
%%%%%%%%%%%%%%%%%%%%\begin{document}
 %%% \maketitle
\section{Introduction}
In this paper we consider the finite automaton based stream cipher published by D\"om\"osi and Horv\'ath \cite{DH}, 
and we show in details that this cipher can be considered as a stream cipher based on quasigroup.  Some additional properties 
are also discussed. The stream cipher in \cite{DH} works, in short, as follows. The cipher consists of a cryptographically 
secure pseudorandom generator and a finite automaton without outputs having the same input and state sets. 
During the encryption the plaintext is read in sequentially character by character. After getting the next (initially the first) 
plaintext character, the system gets simultaneously the next (initially the first)  pseudorandom string of a fixed length which is 
also an input string of the key-automaton. The corresponding ciphertext character will coincide with the state of the key-automaton 
into which this pseudorandom input string takes the automaton from the state which coincides with the corresponding plaintext character. 
The decryption works similarly, using a so-called inverse key-automaton instead of the key automaton such that the input strings will 
be the mirror images of the corresponding pseudorandom strings.

We note that there are several variants of the quasi group based ciphers. Fortunately, we did not find such a solution in the literature that is equivalent to the solution we are discussing. (Detailed overviews and summaries of quasi group based ciphers can be found, for example, in  \cite{CGV,S1, S2}.)
\section{Preliminaries}
We start with some standard concepts and notation. For all notions and notation not defined here we refer to the monographs 
\cite{DN,HMU, MOV, S2} and the reviews \cite{CGV, S1}. 
By an alphabet we mean a finite nonempty set. The elements of an alphabet are called letters. A word over an alphabet $\Sigma$ is 
a finite string consisting of letters of $\Sigma$. The {\it length} of a word $w,$ in
symbols $\vert w\vert,$ means the number of letters in $w$ when each letter
is counted as many times as it occurs.
The string consisting of zero letters is called the
{\it empty word,} written by $\lambda.$
By definition, $\vert\lambda\vert=0.$ The mirror image
$w^R$ of the word $w=a_1\cdots a_n, a_1\ldots, a_n\in \Sigma$ is the word $w^R=a_n\cdots a_1$. By definition, $\lambda^R=\lambda$. 
Furthermore, for every nonempty word $w,$ denote by $\overrightarrow{w}$ the last letter of $w.$
($\overrightarrow{\lambda}$ is not defined.)
The set of all nonempty words over an alphabet $\Sigma$ will be denoted by $\Sigma^+$. In addition, we put $\Sigma^*=\Sigma^+\cup\{\lambda\}$.
By an automaton we mean a finite deterministic automaton without outputs. In other words, by an automaton we mean a system 
${\bf A}=(A,\Sigma,\delta)$ with a finite set $A$ of states, a finite set $\Sigma$ of inputs, and the transition function 
$\delta: A\times\Sigma\to A$. We assume that the transition function of ${\bf A}$ is given in the form of transition table, 
where the lines of this table are denoted by the elements of the input set $\Sigma$ and the columns of this table are denoted by the 
elements of the state set $A$. Therefore, for every input $x$ and state $a$, $\delta(a,x)$ is at the intersection of the row denoted by $x$ and the
column denoted by $a$. 

\section{Automata and Quasigroups}
Given an automaton ${\bf A}=(A,\Sigma,\delta)$,
let $a\in A,a_1,\ldots,a_n\in\Sigma$ and suppose
$b_1=\delta(a,a_1),b_2=\delta(b_1,a_2),\ldots,
\linebreak b_n=\delta(b_{n-1},a_n)$. Then we shall use the notation $\delta(a,a_1\cdots a_n)=b_1\cdots b_n$.
(Thus we may use the notation
$\overrightarrow{\delta(a,a_1\cdots a_n)}$
for the above considered $b_n$.)
Moreover, by definition, $\delta(a,\lambda)=\lambda$.
%%%%%%%%%%%%%%%%!!!!!!!!!!
In what follows we consider automata having the same state and input sets, i.e., we assume $A=\Sigma$. A groupoid $Q=(A,*)$ is a structure  consisting of the nonempty set $A$ and the binary operation $*$ over $A.$
Therefore, the concept of automaton  
${\bf A}=(A,A,\delta)$ coincides with the concept of groupoid
$Q=(A,*)$ having $a*b=\delta(b,a)$ 
%$a*b=\delta(a,b)$ 
for every pair $a,b\in A$. 
A groupoid $Q =(A,*)$ is called a quasigroup if for every pair
$a, b \in A$ there exists unique $x, y \in A$ such that
$a * x = b$ and $y * a = b$. It is easy to see that quasigroups
satisfy both of the  cancellation properties, i.e.,
for every triplet $a,b,c\in A$, $a*b=a*c$ implies $b=c$ (left cancellation), and
$a,b,c\in A$, $a*b=c*b$ implies $a=c$ (right cancellation).
It is said that $\setminus$ is the left inverse operation on $Q$ if for every triplet $a,b,c\in A$,
$a*b=c$ if and only if $b=a\setminus c$. Analogously, $/$ is the right inverse operation on $Q$ if for every triplet $a,b,c\in A$,
$b*a=c$ if and only if $b=c/a$. Then the groupoid $Q_{LI}=(A,\setminus)$ is the left inverse quasigroup of $Q$, and similarly, 
the groupoid  $Q_{RI}=(A,/)$ is the right inverse quasigroup of $Q$. \footnote{It is easy to show that 
	both of $Q_{LI},Q_{RI}$ are quasigroups}

\section{Latin squares and key automata}
A Latin square of order $n$ is an $n\times n$ matrix (with $n$ rows and $n$ columns) in which the elements of an $n$-state set  
$\{a_0,a_1,\ldots,a_{n-1}\}$ are  entered so that each  element  occurs exactly once in each fixed row, and each fixed column, respectively.
We say that ${\bf A}=(A,\Sigma,\delta)$ is a key automaton if for every pair of distinct states $a,b$ and pair of distinct inputs 
$x,y$, $\delta(a,x)$ differs from $\delta(b,x)$ and $\delta(a,x)$ also differs from $\delta(a,y)$. Obviously, in this case 
the transition table of a key automaton forms a Latin square and there is a one-to-one correspondence between the key automata 
${\bf A}=(A,A,\delta)$  and quasigroups $Q=(A,*)$ having the property $\delta(a,b)=b*a$ for every pair $a,b\in A$ of elements 
in $A$  and vice versa.
Given a key automaton ${\bf A}=(A,A,\delta)$ , let us define the automaton ${\bf B}=(A,A,\delta^{-1})$ such that for every pair 
$a,b \in A, \delta^{-1}(\delta(a,b),b)=a$. Then we say that  ${\bf B}$  is the inverse key automaton of the key automaton  ${\bf A}$. 
\medskip

\begin{prop}{\bf 1}
	Every key automaton has exactly one inverse key automaton.	
\end{prop}

\PF
	Consider a key automaton ${\bf A}=(A,A,\delta_{\bf A})$. Suppose that ${\bf B}=(A,A,\delta_{\bf B})$ and
	${\bf C}=(A,A,\delta_{\bf C})$ are inverse key automata of ${\bf A}$ such that ${\bf B}\ne{\bf C}$.
	Then there are $a,b\in A$ having $\delta_{\bf B}(a,b)\ne\delta_{\bf C}(a,b)$. Put 
	$x=\delta_{\bf B}(a,b)$ and $y=\delta_{\bf C}(a,b)$. Then we have $a=\delta_{\bf A}(x,b)=\delta_{\bf A}(y,b)$
	contradicting the assumption that ${\bf A}$ is a key automaton. This completes the proof.
	
\QED
\medskip

\begin{prop}{\bf 2}\label{a}
	Every inverse key automaton is also a key automaton.
\end{prop}

\PF
	Consider a key automaton ${\bf A}=(A,A,\delta)$ and its inverse key automaton  ${\bf B}=(A,A,\delta^{-1})$. 	
	First we suppose that there are states $a,b,c$ with $\delta^{-1}(a,b)=\delta^{-1}(a,c)$ and $b\ne c$. Put
	$d=\delta^{-1}(a,b)=\delta^{-1}(a,c)$.By our assumptions, this implies $\delta(d,b)=\delta(d,c)=a$ with $b\ne c$
	contradicting the assumption that ${\bf A}$ is a key automaton. Thus $b\ne c$ implies $\delta^{-1}(a,b)\ne\delta^{-1}(a,c)$ 
	for every $a\in A$. 	
	Next we suppose that there are states $a,b,c$ with $\delta^{-1}(a,b)=\delta^{-1}(c,b)$ and $a\ne c$. Put
	$d=\delta^{-1}(a,b)=\delta^{-1}(c,b)$.By our assumptions, this implies $\delta(d,b)=a=c$ contradicting to  $b\ne c$.
	Thus $a\ne c$ implies $\delta^{-1}(a,b)\ne\delta^{-1}(c,b)$ 	for every $b\in A$. 
	Therefore, we received that  ${\bf B}$ is also a key automaton.
	
\QED
\medskip

\noindent By the definition of inverse key automaton and Poposition 2
%\ref{a} 
we have as follows.
\medskip

\begin{cor}{\bf 3}\label{b}
	Let ${\bf B}$ be the inverse key automaton of the key automton ${\bf A}$. Then  ${\bf A}$ is the inverse key automaton of  ${\bf B}$.
\end{cor}
\medskip

\begin{prop}{\bf 4}
	Given a key automaton ${\bf A}=(A,A,\delta)$, its inverse key automaton ${\bf A^{-1}}=(A,A,\delta^{-1})$, a state $a\in A$,
	and a string $a_1\cdots a_n, a_1\ldots, a_n\in A$, we have
	$\delta(a,a_1\cdots a_n)=b_1\cdots b_n$ if and only if 
	$\delta^{-1}(b_n,a_n\cdots a_1)\linebreak=b_{n-1}\cdots b_1a$.
\end{prop}

\PF
	Let ${\bf A}=(A,A,\delta)$ be an arbitrary finite automaton and 
	consider (nonempty and finite) strings $a_1\cdots a_n$ and 
	$b_1\cdots b_n$ consisting of the elements $a_1,\ldots,a_n, b_1,\ldots, b_n$ of $A$.	
	In addition, $\delta(a,a_1)=b_1$ if and only if
	$\delta^{-1}(b_1,a_1)=a$, where $\delta^{-1}$ denotes 
	the transition function of the inverse key automaton ${\bf A^{-1}}$ of ${\bf A}$. Similarly,
	$\delta(b_1,a_2)=b_2$ if and only if
	$\delta^{-1}(b_2,a_2)=b_1$. Thus we obtain that 
	$\delta(a,a_1a_2)=b_1b_2$ if and only if 
	$\delta^{-1}(b_2,a_2a_1)=b_1a$. Repeating this procedure we get our statement.
	
\QED
\medskip

We have the following consequence of this statement.
\medskip

\begin{prop}{\bf 5}\label{inv}
	Given a key automaton ${\bf A}=(A,A,\delta)$, its inverse
	key automaton ${\bf A^{-1}}=(A,A,\delta^{-1})$, a state $a\in A$,
	and a string $a_1\cdots a_n, a_1\ldots, a_n\in A$, we have
	$\overrightarrow{\delta(a,a_1\cdots a_n)}=b_n$ if and only if 
	$\overrightarrow{\delta^{-1}(b_n,a_n\cdots a_1)}=a$.
\end{prop}
%\medskip

\section{Quasigroups}

We shall use the following statement.
\medskip

\begin{prop}{\bf 6}\label{QLI}
	Given a quasigroup $Q =(A,*)$, its left inverse quasigroup $Q_{LI}=(A,\setminus)$, 
	moreover, $a_1,\ldots,\linebreak a_n, a,b\in A$. Then
	$a_n*(\cdots *(a_2*(a_1*a))\cdots)=b$ if and only if $a_1\setminus(\cdots \setminus (a_{n-1}\setminus(a_n\setminus b))\cdots)=a$.
\end{prop}

\PF
	We will prove our statement by induction. 
	Suppose $n=1$. Then, by definition, $b=a_1*a$ if and
	only if $a=a_1\setminus b$. Thus, it is enough to show than if our statement holds for any
	 given case $n=m$, then it must also hold for the next case $n=m+1$.  
 Thus, assume that for every $b,c,a_2,\ldots,a_{m+1}\in A$,  $a_{m+1}*(\cdots *(a_2*c)\cdots)=b$
	if and only if $a_{2}\setminus(\cdots \setminus (a_{m+1}\setminus b))\cdots)=c$.
	Set $c=a_1*a$ for some $a_1\in A$.
	Then $a_{m+1}*(\cdots *(a_2*(a_1*a))\cdots)=b$
	if and only if $a_{2}\setminus(\cdots \setminus (a_{m+1}*b))\cdots)=a_1*a$.
	
	Substituting $a_{2}\setminus(\cdots \setminus (a_{m+1}*b))\cdots)$ for $b$, then we receive 
	$b=a_1*a$ which follows $a=a_1\setminus b$ by definition. This implies 
	$a_1\setminus a_{2}\setminus(\cdots \setminus (a_{m+1}*b))\cdots)=b$ as we stated.

\QED
\medskip

The following statement is obvious.
\medskip

\begin{prop}{\bf 7}\label{left}
	Given a quasigroup $Q=(A,*)$, let 
	$Q_{LI}=(A,\setminus)$ be its left inverse quasigroup.
	Then for every pair $x,y\in A$,
	$x\setminus (x * y) = y, x * (x\setminus y) = y$.
\end{prop}
\medskip
Given  a key automaton ${\bf A}=(A,A,\delta)$, the corresponding quasigroup $Q=(A,*)$ ordered to $\cal A$ is defined by $a*b=\delta(a,b), a,b\in A$. 

\medskip

\begin{theo}{\bf 8}\label{QA}
	Let $Q=(A,*)$  be the corresponding quasigroup ordered to the key automaton ${\bf A}=(A,A,\delta)$. Then the left
	inverse quasigroup   of $Q$ is the corresponding quasigroup ordered to the inverse key automaton of ${\bf A}=(A,A,\delta)$ and
	vice versa.
\end{theo}

\PF
	Consider a key automaton ${\bf A}=(A,A,\delta)$ and its inverse key automaton  ${\bf B}=(A,A,\delta^{-1})$.

	Then the corresponding quasigroup $Q=(A,*)$ ordered to ${\bf A}$ has the property $\delta(a,b)=b*a$ for every pair $a,b\in A$.
	Similarly, the corresponding quasigroup $R=(A,\bigodot)$ ordered to ${\bf B}$ has the property 
	$\delta^{-1}(c,d)=d\bigodot c$ for every pair $c,d\in A$. 
	
	%$a*b=c$ if and only if $b=a\setminus c$.
	
	By definition, for every pair $a,b \in A, \delta^{-1}(\delta(a,b),b)=a$. This implies
	$b\bigodot(b*a)=a$. Then $c=b*a$ implies $b\bigodot c =a$. 
	
	Next we assume $b\bigodot(b*a)=a$ and $c\ne b*a$ with $b\bigodot c=a$.
	Put $d=b*a$. Then we get $b\bigodot d=a$ with $b\bigodot c=a$ and $d\ne c$.
	In other words,  $\delta^{-1}(d,b)=\delta^{-1}(c,b) (=a)$ with $d\ne c$.
	But then, by definition,  the inverse key automaton ${\bf B}$ is not a key automaton. This statement contradicts to Proposition 2. %\ref{a}.
	
\QED
\medskip

\begin{prop}{\bf 9}\label{main} 
	Given a quasigroup $Q=(A,*)$ ordered to the key automaton 
	${\bf A}=(A,A,\delta)$,let $a, a_1,\ldots, a_n\in A$. 
	Then $\overrightarrow{\delta(a,a_1\cdots a_n)}=b$ 
	for some $b\in A$ if and only if
	$a_n*(a_{n-1}*(\cdots *(a_1*a)\cdots))=b$.  
\end{prop}

\PF
	By our conditions, we have in order, 
	$\overrightarrow{\delta(a,a_1)}=\delta(a,a_1)= a_1*a$,
	$\overrightarrow{\delta(a,a_1a_2)}=\linebreak\overrightarrow{\delta(a,a_1)
		\delta(\delta(a,a_1),a_2)}=\delta(\delta(a,a_1),a_2)=
	a_2*(a_1*a)$,
	and inductively, 
	$\overrightarrow{\delta(a,a_1\cdots a_n)}= 
	a_n*(a_{n-1}*(\cdots *(a_1*a)\cdots))$.
	Using these observations, 
	by definition, $b_1=\delta(a,a_1)$ if and only if $b_1=a_1*a$.
	Similarly, $b_2=\delta(b_1,a_2)$ if and only if $b_2=a_2*(a_1*a)$.
	Repeating this procedure, we have 
	$b_n=\delta(b_{n-1},a_n)$ if and only if $b_n=a_n*(a_{n-1}*(\cdots *(a_1*a)\cdots))$.
	
	Let $b=b_n$. Then we get as we stated.
	
\QED

\section{A finite automaton based stream cipher}

Consider a %%cryptographically secure 
pseudorandom number generator, a key automaton ${\bf A}=(A,A,\delta)$, and its inverse key automaton 
${\bf A^{-1}}=(A,A,\delta^{-1})$. The main idea of the discussed cipher is the following.

%We extend $\delta: A\to A$ to $\delta: A^+\to A$ in %the following way. Consider a string  

\subsection{Encryption}

Let $p_1\cdots p_n, p_1,\ldots,p_n\in A$ be a plaintext and let $r_1,\ldots,r_n\in A^+$  be pseudorandom strings  of the same fixed length $m\geq 1$ 
generated by a given %cryptographically secure
 pseudorandom number generator starting by a seed  $r_0$. 
We note that $\vert r_0\vert,\ldots, \vert r_k\vert = m$ holds for a fixed positive integer $m$. 

The ciphertext will be $c_1\cdots c_n, c_1,\ldots, c_n\in A$ with $c_1=\overrightarrow{\delta(p_1,r_1)},\ldots, c_n=\overrightarrow{\delta(p_n,r_n)}$.

\subsection{Decryption}

Let $c_1\cdots c_n, c_1,\ldots, c_n\in A$ be a ciphertext and let $r_1,\ldots,r_n\in\Sigma^+$  be the same pseudorandom strings generated by the pseudorandom number generator starting by a seed  $r_0$. 

The decrypted plaintext  will be $p_1\cdots p_n$ with $p_1=\overrightarrow{\delta^{-1}(c_1,(r_1)^R)},\ldots, p_n=\overrightarrow{\delta^{-1}(c_n,(r_n)^R)}$.

The next statement shows the correctness of the discussed finite
automaton-based encryption and decryption procedure. 
\medskip

\begin{theo}{\bf 10}\label{Q} Let $p_1\cdots p_n, p_1,\ldots,p_n\in A$ be a plaintext and let $r_1,\ldots,r_n\in A^+$  be pseudorandom strings of the same fixed length $m\geq 1$ generated by a given
	%cryptographically secure 
	pseudorandom number generator starting by a seed  $r_0$.
	Moreover, let ${\bf A}=(A,A,\delta)$ be a key automaton and let 
	${\bf A^{-1}}=(A,A,\delta^{-1})$ be its inverse key automaton. 
	If $c_1\cdots c_n$ is the ciphertext generated  by the above finite automaton  encryption procedure then $p_1\cdots p_n$ is the only plaintext
	which can be generated by the above finite automaton based decryption procedure
	(assuming that the pseudorandom generator of the cipher generates the same sequence $r_1,\ldots,r_n$ of the pseudorandom strings during the
	encryption and also during the decryption).
\end{theo}

\PF
	Consider a key automaton ${\bf A}=(A,A,\delta)$, its inverse key automaton ${\bf A^{-1}}=(A,A,\delta^{-1})$, a state $a\in A$,
	and a string $r\in A$. By Proposition 5 %\ref{inv} 
	we have
	$\overrightarrow{\delta(a,r)}=b_n$ if and only if 
	$\overrightarrow{\delta^{-1}(b_n,r^R)}=a$.\footnote{Recall that for every $r\in A^+, r^R$ denotes the mirror image of $r$.}
	
	By our construction, for every $i=1,\ldots,n,
	c_i=\overrightarrow{\delta(p_i,r_i)}$. By Proposition 5%\ref{inv} this is possible if and only if 
	$p_i=\overrightarrow{\delta^{-1}(c_i,r_i^R)}$. 
	In sum, $c_1\cdots c_n=\overrightarrow{\delta(p_1,r_1)}\cdots
	\overrightarrow{\delta(p_n,r_n)}$ which, by Proposition 5 %\ref{inv}
	is possible if and only if 
	$p_1\cdots p_=\linebreak\overrightarrow{\delta^{-1}(c_1,r_1^R)}\cdots
	\overrightarrow{\delta^{-1}(c_n,r_n^R)}$. This completes the proof.
	
\QED

\section{A quasigroup based stream cipher}

Consider again  a cryptographically secure pseudorandom number generator, moreover a quasigroup\linebreak $Q=(A,*)$ and its left-inverse $Q_{LI}=(A,\setminus)$. The main idea of the discussed cipher is the following. 

\subsection{Encryption}\label{EQ}

Let $m$ be a fixed positive integer, and in order to have $P_T = p_1p_2p_3\cdots p_n,\ p_1,\ldots,p_n\in A$, $K=k_{1,1}k_{1,2}\cdots\linebreak k_{1,m}\cdots k_{n,1}k_{n,2}\cdots k_{n,m}$, $C_T=c_1c_2\cdots c_n$ as the plaintext $P_T$ to be encrypted, a pseudorandom sequence $K$  is generated by the cryptographically secure pseudorandom  number generator as the keystream to be used for encryption, and the resulting ciphertext $C_T$ respectively. Then a way of encrypting $P_T$ with the keystream $K$ to obtain the corresponding $C_T$ is as follows:  
%$b_n=a_n*(a_{n-1}*(\cdots *(a_1*a)\cdots))$.

$c_1=k_{1,m}*(\cdots *(k_{1,2}*(k_{1,1}*p_1))\ldots), c_2=k_{2,m}*(\cdots *(k_{2,2}*(k_{2,1}*p_2))\ldots),\ldots
c_n=k_{n,m}*(\cdots *(k_{n,2}*(k_{n,1}*p_n))\ldots)$.

\subsection{Decryption}

Let $m$ be the same fixed positive integer again as in Subsection 6.1, %\ref{EQ}, 
and in order to have the same $C_T=c_1c_2\cdots c_n$, $K=k_{1,1}k_{1,2}\cdots k_{1,m}\cdots k_{n,1}k_{n,2}\cdots k_{n,m}$ as in Section 6.1, %\ref{EQ}, 
as the ciphertext to be decrypted,  a pseudorandom sequence $K$  is generated by the cryptographically secure pseudorandom  number generator as the keystream to be used for encryption,and the resulting plaintext $P_T = p_1p_2p_3\cdots p_n$, respectively. Then a way of decrypting $C_T$ with the keystream $K$ to obtain the corresponding $P_T$ back is as follows:  
$p_1=k_{1,1}\setminus(\cdots\setminus(k_{1,m-1}\setminus
(k_{1,m}\setminus c_1))\ldots), 
p_2=k_{2,1}\setminus(\cdots\setminus(k_{2,m-1}\setminus( k_{2,m}\setminus c_2))\ldots) ,
p_n=k_{n,1}\setminus(\cdots\setminus(k_{n,m-1}\setminus k_{n,m}\setminus c_n))\ldots)$, 
where $\setminus$ denotes the  quasigroup operation and  $\setminus$  denotes the corresponding left inverse quasigroup operation.

Next we show that
%, by the above Proposition, 
the work of the discussed stream cipher can be written easily by using automata-theoretic disciplines like in \cite{DH}. 
%More exactly, by Theorem 10 %\ref{Q} 
%we have as follows. 
In more details, the next statement shows the correctness of the discussed finite
qusigroup based encryption and decryption procedure. 

%%%%%%%%%%%%%%%%%%%%%%%%%%%%%%%%%%EZT megnézni:

\begin{theo}{\bf 11}\label{R} Let $p_1\cdots p_n,\ p_1,\ldots, p_n\in A$ be a plaintext and let $r_1,\ldots,r_n\in A^+$  be random strings of the same fixed length $m\geq 1$ generated by a cryptographically secure pseudorandom number generator starting by a seed  $r_0$.
	Moreover, let $Q=(A,*)$ be a quasigroup and let 
	$Q_{LI}=(A,\setminus)$ be its left inverse quasigroup. 
	If $c_1\cdots c_n$ is the ciphertext generated  by the above qusigroup based encryption procedure then $p_1\cdots p_n$ is the only plaintext
	which can be generated by the above qusigroup based  decription procedure
	(assuming that the pseudorandom generator of the cipher generates the same sequence $r_1,\ldots,r_n$ of the pseudorandom strings during the
	encryption and also during the decryption).
	{\tiny }	\end{theo}

\PF
	By Proposition 6,%\ref{QLI},
	\ for every quasigroup $Q =(A,*)$, its left inverse quasigroup $Q_{LI}=(A,\setminus)$, 
	and $a_1,\ldots,a_n,a,b\in A$ it holds that
	$a_n*(\cdots *(a_2*(a_1*a))\cdots)=b$ if and only if $a_1\setminus(\cdots \setminus (a_{n-1}\setminus(a_n\setminus b))\cdots)=a$.
	Let $a$ denote the $i^{th}$ character $p_i$ of the plaintext, moreover, let $b$ denote the $i^{th}$ character $c_i$ of the
	ciphertext for some $i\in\{1,\ldots,n\}$. In addition, let $a_1\cdots a_m$ denote the $i^{th}$ pseudorandom string $r_i$ generated by the
	pseudorandom generator of the cipher. Then, by Proposition 6, %\ref{QLI}, 
	we have that for every $i=1,\ldots,n$, that $p_i$ is the only
	$i^{th}$ plaintext character which can be generated by the discussed qusigroup based decryption procedure whenever $c_i$ is the 
	$i^{th}$ ciphertext character which can be generated by the discussed qusigroup based encryption procedure and $r_i$ is the same 
	$i^{th}$ pseudorandom string generated by the pseudorandom generaton in both of the encryption and the decryption.
	Therefore, if  $c_1\cdots c_n$ is the ciphertext generated  by the considered qusigroup based encryption procedure then $p_1\cdots p_n$ is the only plaintext
	which can be generated by the considered qusigroup based  decription procedure
	(assuming that the pseudorandom generator of the cipher generates the same sequence $r_1,\ldots,r_n$ of the pseudorandom strings during the
	encryption and also during the decryption).This completes the proof. 
\QED
%%%%%%%%%%%%%%%%%%%%%%%%%%%%%%%%

\section{Quasigroups in Cryptography}

The most of the quasigroup-based cryptosystems essentially work based on the following principle \cite{CGV, S1}.
\medskip

Given a quasigroup $Q = (A, *)$, its left inverse quasigroup $Q_{LI} = (A,\setminus)$, let  $\ell\in A$ be a fixed
element, which is called a leader. (Actually, $\ell\in A$ can be considered as the secret seed of the encryption/decryption).
\medskip

{\bf Encryption.} Let $p_1 \cdots  p_n$ be a plaintext of $n\geq 1$ letters $p_1 ,\ldots,  p_n\in A $ . Compute $c_1 = \ell* p_1, c_2 = c_1 * p_2,\ldots, c_n = c_{n-1}*p_n$. 
Then the ciphertext  is $c_1 \cdots c_n$. 
\medskip

{\bf Decryption} Let $c_1\cdots  c_n$ be a plaintext of $n$ letters $c_1 ,\ldots,  c_n \in A$ . Compute $p_1 = \ell\setminus c_1,  p_2 = c_1\setminus c_2, \ldots , p_n = c_{n-1} \setminus c_n$.
Then the recovered plaintext is $p_1 \cdots p_n$ .
\medskip

Cryptanalyses of this classical quasigroup-based cipher was made by M. Vojvoda \cite{VO}. He showed that this cipher 
is not resistant to chosen plaintext attack and ciphertext-only attack in contrast to our discussed solution. 
There are several known variants of this classical quasigroup cipher applying special quasigroups, and/or 
multiple leaders, multi-round ciphering, etc.  \cite{CGV}.

%\section{Automata and Quasigroups}
\section{Conclusion}

This paper shows that the cipher in \cite{DH} can be considered as a quasigroup-based stream cipher. 
By this observation, we can easily compare it with the other quasigroup-based ciphers. It can be concluded that our solution
is mainly different from them. 

%This paper is devoted to propose a novel cryptosystem based on finite automata.

In order, to achieve a higher speed of encryption/description operation, $m$ should be as small as possible. Therefore, next we should analyse this cipher with $m=1$.
%, where the considered pseudorandom number generator is cryptographically secure.
Thus, using the finite automaton-based form,  we should consider again a 
%cryptographically secure pseudorandom generator, a quasigroup and its left-inverse. Let $P_T = p_1p_2p_3\cdots p_n, K=k_1k_2\cdots k_n, C_T=c_1c_2\cdots c_n$  be the plaintext to be encrypted, the pseudorandom sequence generated by the pseudorandom  number generator as the keystream to be used for encryption, and the resulting ciphertext, respectively. Then a way of encrypting $P_T$ with the keystream $K$ to obtain the corresponding $CT$ is as follows: 
$c_1=\delta(p_1,k_1),\ldots, c_n=\delta(p_n,k_n)$, and the description  can  be given by  $p_1=\delta^{-1}(c_1, k_1), p_2=\delta^{-1}(c_2, k_2),\ldots, p_n=\delta^{-1}(c_n, k_n)$,  where   $\delta$ denotes the transition function of the key automaton and  $\delta^{-1}$  denotes the transition function of the inverse key automaton.  

The equivalent quasigroup-based form of this cipher can also be considered as follows:\linebreak
$c_1=p_1*k_1, c_2=p_2*k_2,\ldots, c_n= p_n*k_n$,  and the description  can  be given by  $p_1=c_1\setminus k_1,\linebreak  p_2=c_2\setminus k_2,\ldots, p_n=c_n\setminus k_n$,  where   $*$ denotes the  quasigroup operation and  $\setminus$  denotes the corresponding right inverse quasigroup operation.

A further challenge of research is to show  the security of the proposed cipher using several theoretical and experimental investigations regarding
the length of the applied pseudorandom sequences, the number of rounds in multi-round encryption and decryption,  and some other parameters. 
%Finally, we note that a proposed implementation, its security, and its performance ratio is %described in \cite{DDH}. Therefore, in this paper we omit them.  

% and Future Work}

%%%%%%%%%%%%//nd{figure*}

%\subsection*{Acknowledgement}
%This 
%	work was supported by United Arab Emirates Program for Advanced Research (UAEU UPAR)  Grant \# G00003431.

% bibliography style is not predefined; authors can apply their own,
% but it should be sent to matjaz.gams@ijs.si with the accepted paper

\bibliographystyle{eptcs}
\bibliography{generic}

\end{document}